\documentclass[a4paper]{jpconf}  
\usepackage[dvips]{graphicx}
\usepackage{dcolumn}   
\usepackage{bm}        
\usepackage{amssymb}   
\usepackage[utf8]{inputenc}
\usepackage{mathrsfs}
\usepackage{amsmath}
\usepackage[breaklinks=true,colorlinks=true]{hyperref}
\hypersetup{colorlinks=true,citecolor=blue,linkcolor=blue,urlcolor=blue}
\usepackage{textcomp}
\usepackage{subfigure}
\usepackage{color}

\begin{document}

\title{Asymmetric kink solutions of hyperbolically deformed model}

\author{Vakhid A. Gani$^{1,2}$ and Aliakbar Moradi Marjaneh$^3$}

\address{$^1$Department of Mathematics, National Research Nuclear University MEPhI\\ (Moscow Engineering Physics Institute), Moscow 115409, Russia}

\address{$^2$Theory Department, Institute for Theoretical and Experimental Physics\\ of National Research Centre ``Kurchatov Institute'', Moscow 117218, Russia}

\address{$^3$Department of Physics, Quchan Branch, Islamic Azad university, Quchan, Iran}

\ead{vagani@mephi.ru}

\begin{abstract}
We study some properties of kink solutions of the model with non-polynomial potential obtained by deforming the well-known $\varphi^6$ field model. We consider the excitation spectrum of the kink. We also discuss the properties of the `kink+antikink' system as a whole that are not inherent to a solitary kink or antikink.
\end{abstract}

\section{Introduction}

The properties of kinks of the $\varphi^6$ model have been fairly well studied. In particular, the interactions between the kink and the antikink were investigated both using the collective coordinate method and by numerically solving the equation of motion \cite{GaKuLi,Demirkaya.JHEP.2017,Weigel.JPCS.2014}. Interesting results have been obtained on various phenomena in the kink-antikink and multi-kink scattering \cite{MGSDJ.JHEP.2017,Dorey.PRL.2011,Romanczukiewicz.PLB.2017,Lima.JHEP.2019}.

On the other hand, there is the so-called deformation procedure \cite{Bazeia.PRD.2002,Bazeia.PRD.2004,Bazeia.EPJC.2018,Gani.arXiv.2020.deformations}, which allows to obtain a new model along with its kink solution from known model with its known kink solution.

In this paper, we apply the deformation procedure to the $\varphi^6$ model, using the hyperbolic sine as the deforming function. The result of such deformation is the sinh-deformed $\varphi^6$ model. (The fact that the study of the properties of kinks of the sinh-deformed $\varphi^6$ model is of interest was indicated in \cite{Bazeia.EPJC.2018}.) The preliminary results reported by Dr.\ Aliakbar Moradi Marjaneh at the ICPPA-2020 conference indicate that in collisions of the kink and antikink of the sinh-deformed $\varphi^6$ model resonance phenomena (escape windows) are present; see, e.g., \cite{Belova.UFN.1997} for details about escape windows. At first glance, this contradicts the fact that there are no vibrational modes in the excitation spectrum of the kink. However, if we recall papers \cite{Dorey.PRL.2011,Belendryasova.CNSNS.2019}, then we understand that everything is not so simple.

So, let's move on to a slightly more detailed presentation of our idea. Emphasize that this text cannot be regarded as a complete study, but only as a brief presentation of preliminary results.

\section{The $\varphi^6$ model and its hyperbolic deformation}

Within the $\varphi^6$ model the dynamics of a real scalar field $\varphi(x,t)$ is described by the Lagrangian density
\begin{equation}\label{eq:Largangian}
	\mathscr{L} = \frac{1}{2} \left( \frac{\partial\varphi}{\partial t} \right)^2 - \frac{1}{2} \left( \frac{\partial\varphi}{\partial x} \right)^2 - V(\varphi)
\end{equation}
with the potential
\begin{equation}\label{eq:potential}
    V^{(0)}(\varphi) = \frac{1}{2}\varphi^2\left(1-\varphi^2\right)^2.
\end{equation}
This model has kink solutions (two kinks and two antikinks) connecting the vacua $\varphi=0$ and $\varphi=\pm 1$. All of them can be obtained by symmetry transformations from the kink
\begin{equation}\label{eq:phi6_kink}
\varphi_{\rm K}^{(0)}(x) = \sqrt{\frac{1+\tanh x}{2}},
\end{equation}
see, e.g., \cite{GaKuLi}.

We apply the deformation procedure to the $\varphi^6$ model, using the deforming function $f(\varphi)=\sinh\varphi$. As a result, we obtain the sinh-deformed $\varphi^6$ model with the potential
\begin{equation}\label{eq:sinh_deformed_phi6_potential}
V^{(1)}(\varphi)=\frac{1}{2} \tanh^2 \varphi(1-\sinh^2 \varphi)^2
\end{equation}
and the kink solution, corresponding to the $\varphi^6$ kink \eqref{eq:phi6_kink}:
\begin{equation}\label{eq:deformed_phi6_kink}
\varphi_{\rm K}^{(1)}(x) = \mbox{arsinh}\:\sqrt{\frac{1+\tanh x}{2}}.
\end{equation}
In the next section, we will discuss the stability potential of the sinh-deformed $\varphi^6$ kink \eqref{eq:deformed_phi6_kink}.

%
%

%
%

\section{Stability potential of the sinh-deformed $\varphi^6$ kink}

First of all, the stability potential of a kink can be obtained by adding a small perturbation to the static kink. Then the linearized equation of motion leads to the Sturm-Liouville problem
\begin{equation}\label{eq:stat_Schr}
\left[-\frac{d^2}{dx^2} + U(x)\right]\psi(x) = \omega^2\psi(x),
\end{equation}
see, e.g., \cite[Sec.~2]{Bazeia.EPJC.2018} for more details. The function
\begin{equation}\label{eq:Schr_pot}
U(x) = \left.\frac{d^2 V^{(1)}}{d\varphi^2}\right|_{\varphi_{\rm K}^{(1)}(x)}=\frac{30 \tanh x + 2\:\text{sech}^4 x - \left(16\tanh x + 45\right) \text{sech}^2x + 34}{\left(\tanh x + 3\right)^2}
\end{equation}
is the stability potential, which defines the kink's excitation spectrum. The structure of the discrete part of the spectrum is very important for various processes with kinks. The spectrum always has zero level, see, e.g., \cite[Eqs.~(25), (26)]{Belendryasova.CNSNS.2019}, and all eigenvalues of the problem \eqref{eq:stat_Schr} are non-negative \cite[Eqs.~(2.20), (2.21)]{Bazeia.EPJC.2018}.

\begin{figure}[t!]
\begin{center}
  \centering
  \subfigure[two-bounce escape window, the initial velocity is 0.0456]{\includegraphics[width=0.49\textwidth]{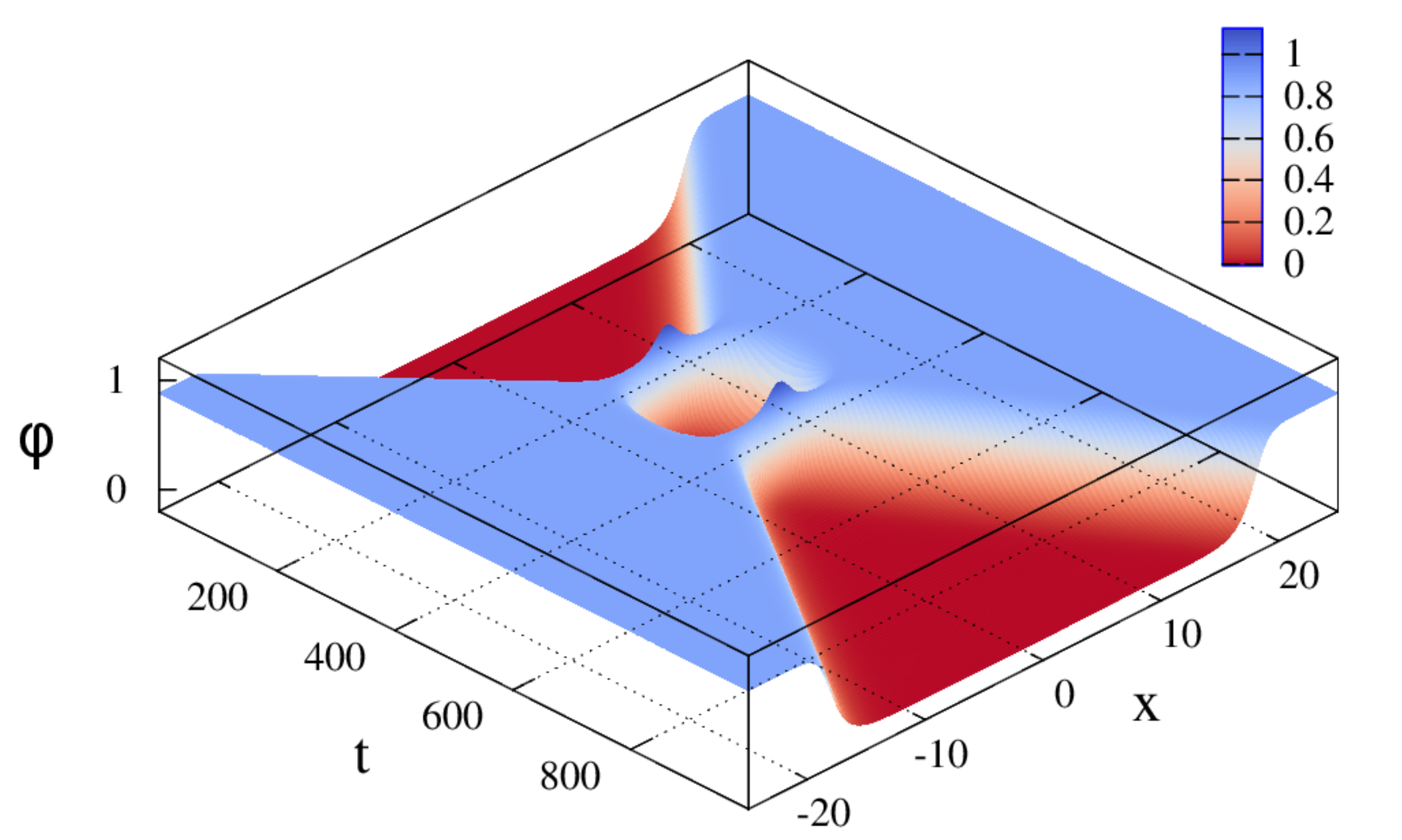}\label{fig:phi_a0a_00456}}
    \subfigure[three-bounce escape window, the initial velocity is 0.0441]{\includegraphics[width=0.49\textwidth]{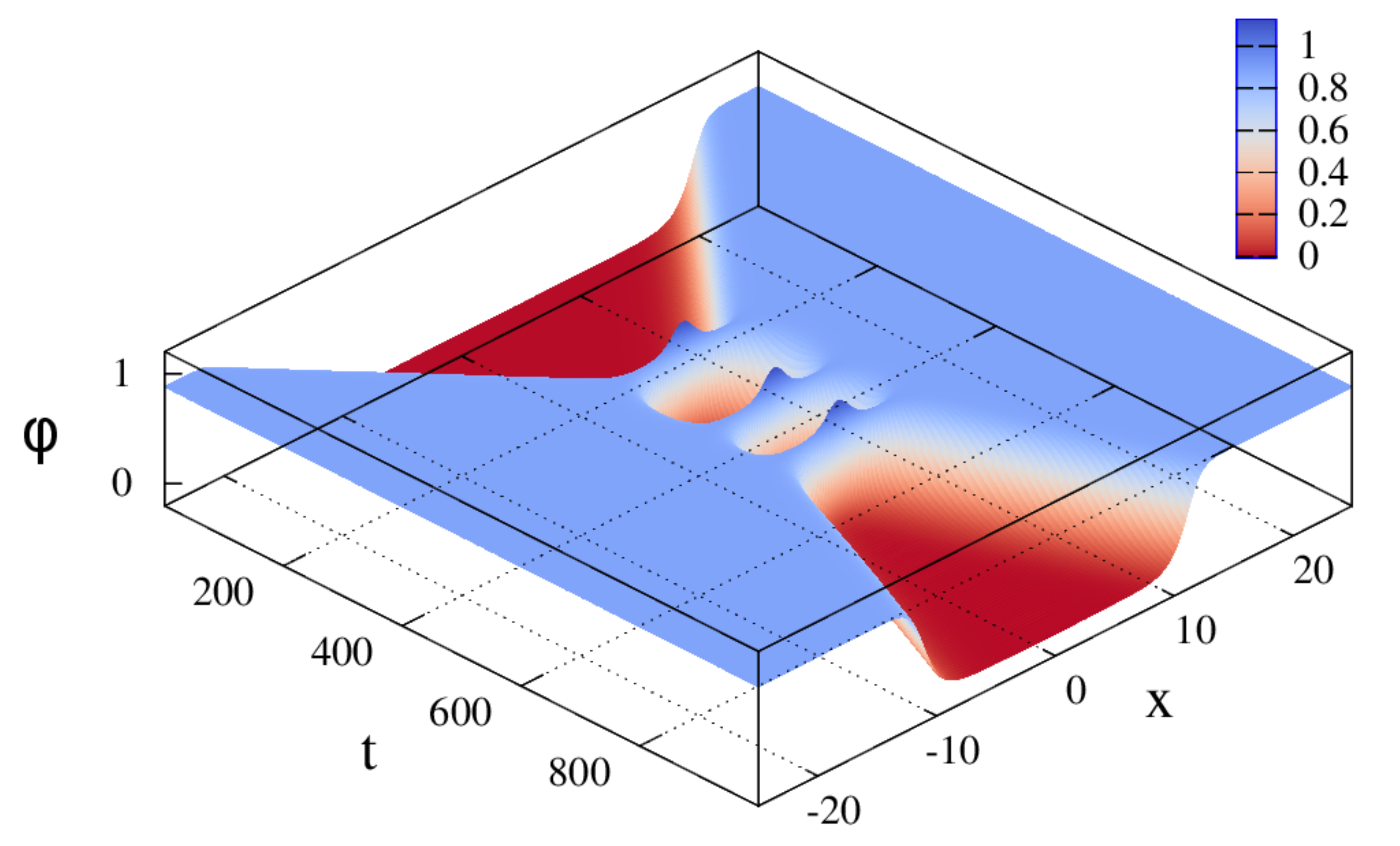}\label{fig:phi_a0a_00441}}
  \caption{Examples of escape windows in the antikink-kink scattering. The initial positions of the kinks are $\pm 20$.}
  \label{fig:a0a_bounce_collision}
\end{center}
\end{figure}

It is known that the stability potential of the $\varphi^6$ kink does not have vibrational modes. We found that the stability potential \eqref{eq:Schr_pot} of the sinh-deformed $\varphi^6$ kink also does not have vibrational modes. It would seem that it can be assumed that resonance phenomena associated with resonant energy exchange between zero and vibrational modes are impossible in the kink-antikink collisions in both models. However, this is not quite true. In the sinh-deformed $\varphi^6$ antikink-kink collisions we observe escape windows, see Fig.~\ref{fig:a0a_bounce_collision}.
Appearance of the escape windows indicates the presence of resonant energy exchange between kinetic energy and (at least one) vibrational mode. The question is where is this vibrational mode hiding?

\begin{figure}[t!]
\begin{center}
  \centering
  \subfigure[]{\includegraphics[width=0.49\textwidth]{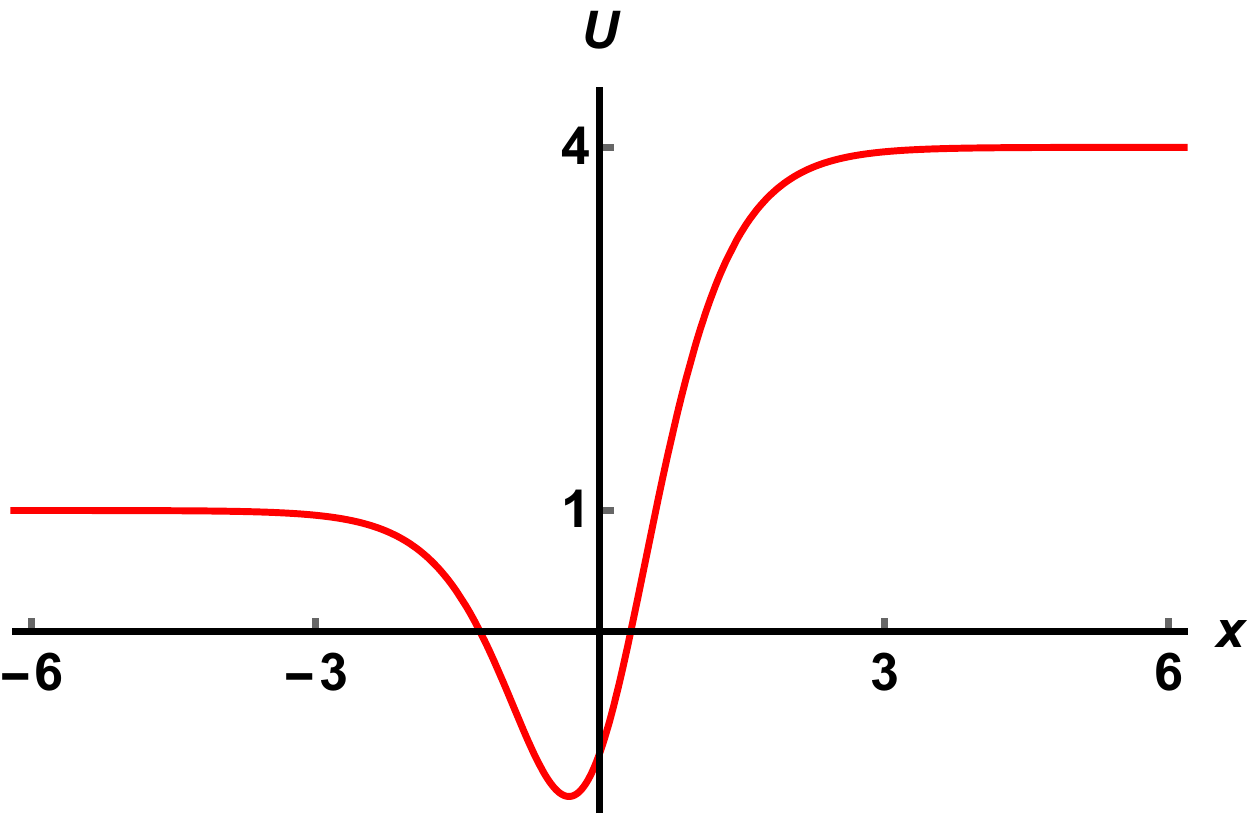}\label{fig:QMPkink}}
    \subfigure[]{\includegraphics[width=0.49\textwidth]{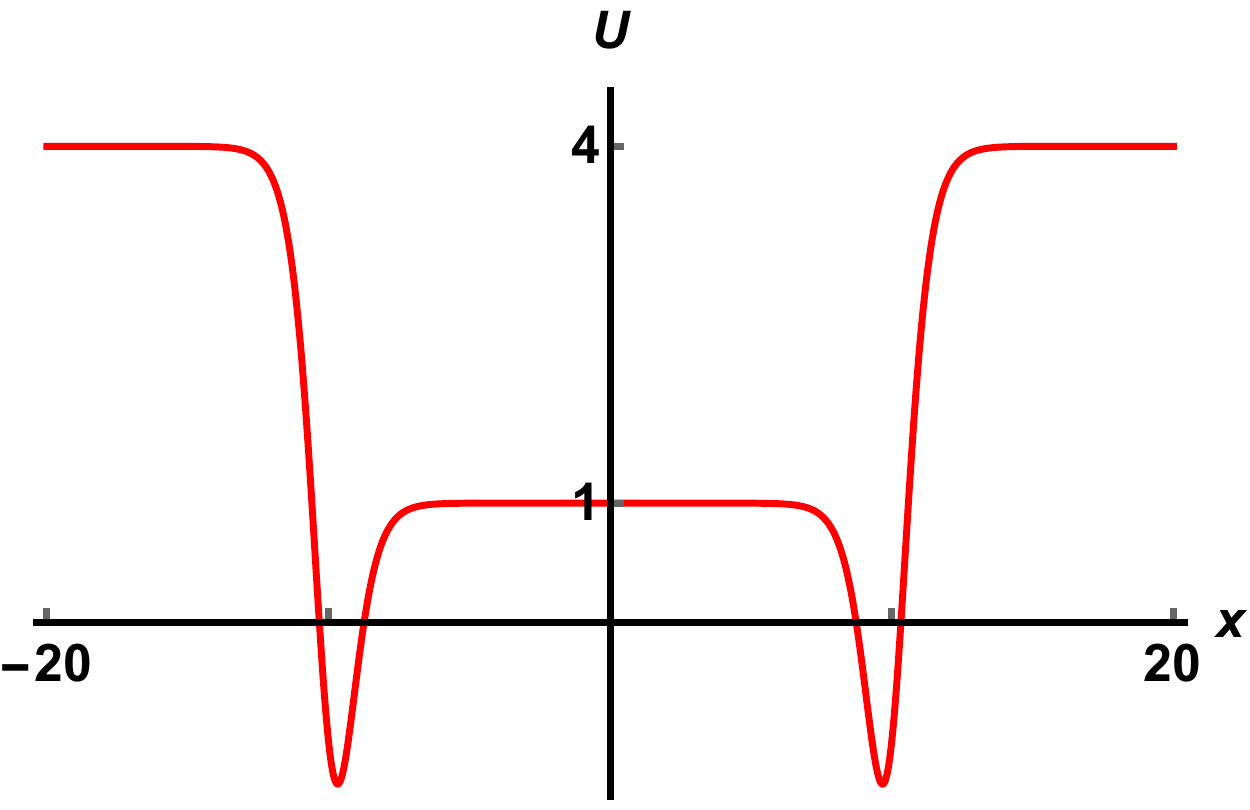}\label{fig:QMPkinkantikink}}
  \caption{Stability potential for (a) kink and (b) `antikink+kink' configuration with kink and antikink at $x=\pm 10$, i.e.\ $x_0^{}=10$ in Eq.~\eqref{eq:Schr_pot_kk}.}
  \label{fig:QMP}
\end{center}
\end{figure}

The answer can be in that the kinks of both models are asymmetric. This leads to the asymmetric stability potentials with different asymptotic values at $x\to\pm\infty$, see Fig.~\ref{fig:QMPkink}.
Such asymmetry means that closely placed kink and antikink can form a mutual stability potential
\begin{equation}\label{eq:Schr_pot_kk}
U(x) = \left.\frac{d^2 V^{(1)}}{d\varphi^2}\right|_{\varphi_{\rm \bar K}^{(1)}(x+x_0^{})+\varphi_{\rm K}^{(1)}(x-x_0^{})}
\end{equation}
in the form of a potential well, Fig.~\ref{fig:QMPkinkantikink}, in which, in addition to the zero level, there will also be levels of the discrete spectrum (vibrational modes). (For the sake of convenience, in Fig.~\ref{fig:QMPkinkantikink} we used $x_0^{}=10$, which is not small.) Our idea is that the situation can be similar to that observed for the $\varphi^6$ kinks in \cite{Dorey.PRL.2011} or for the $\varphi^8$ kinks in \cite{Belendryasova.CNSNS.2019}.

Note that preliminary results show presence of vibrational modes in the potential well of Fig.~\ref{fig:QMPkinkantikink}.

\section{Conclusion}

We have studied the sinh-deformed $\varphi^6$ model which is obtained from the well-known $\varphi^6$ scalar field theory. We have shown that in this new model there are no vibrational modes in the kink excitation spectrum. At the same time, in our numerical simulations of collisions of the sinh-deformed $\varphi^6$ kink and antikink at some initial velocities we observed resonance phenomena --- escape windows. We suppose that these resonance phenomena may be a consequence of the resonant energy exchange between the translational modes of kinks (their kinetic energy) and the vibrational modes of the `antikink+kink' system as a whole. A more detailed study of this issue is of great importance and is planned for the near future.

\section*{Acknowledgements}

V.A.G.\ acknowledges the support of the Russian Foundation for Basic Research under Grant No.\ 19-02-00930.

A.M.M.\ thanks the Islamic Azad University, Quchan Branch, Iran (IAUQ) for their financial support under the Grant.

The work of the MEPhI group was supported by the MEPhI Academic Excellence Project.

\end{document}